\documentstyle[12pt]{article}
\begin{document}

\title{A quantum phase gate implementation for trapped ions in thermal motion}
\author{XuBo Zou, K. Pahlke and W. Mathis\\
Electromagnetic Theory Group at THT,\\
 Department of Electrical
Engineering, \\
University of Hannover, Germany}

\date{\today}
\maketitle
\begin{abstract}
We propose a novel scheme to implement a quantum controlled phase
gate for trapped ions in thermal motion with one standing wave
laser pulse. Instead of applying the rotating wave approximation
this scheme makes use of the counter-rotating terms of operators.
We also demonstrate that the same scheme can be used to generate
maximally entangled states of $N$ trapped ions by a single laser
pulse. 03.67.Lx, 03.65.Ud ,42.50.Dv.
\end{abstract}
The existence of quantum algorithms for specific problems shows
that a quantum computer can in principle provide a tremendous
speed up compared to classical computers \cite{shor,gro}. This
discovery motivated an intensive research into this mathematical
concept which is based on quantum logic operations on multi-qubit
systems \cite{bare}. In order to implement this concept into a
real physical system a quantum system is needed, which makes the
storage and the read out of quantum information and the
implementation of the required set of quantum gates possible. This
system should be scalable and the isolation of the system from the
environment should be very well in order to suppress decoherence
processes. Several physical systems were suggested to implement
the concept of quantum computing: cavity QED systems \cite{tc},
trapped ion systems \cite{cd}, nuclear magnetic resonance systems
\cite{cg}, solid state systems with nuclear spins \cite{ka},
quantum dot systems \cite{loss} and Josephson-junction device
systems \cite{mak}. The ion trap quantum information processor,
first proposed by Cirac and Zoller \cite{cz}, was demonstrated to
be a promising candidate for the realization of a small scale
quantum computations \cite{cm}. The original idea of Cirac and
Zoller uses one mode of the collective ion vibration for a bus
concept. The implementation of the Cirac/Zoller-gate requires to
cool the mode of the collective vibration to the ground state. Any
heating process on this mode, which is caused by the environment,
will diminish the accuracy of the Cirac/Zoller-gate. Thus, an
unreliable performance of the quantum computer as a whole will
result, which makes the implementation of even simple algorithms
impossible. The elimination of heating processes is a very
demanding goal but not possible. Several gate proposal do not pose
the requirement of ground state cooling \cite{pcz,sm,sjj,gjm}. In
this paper we propose a novel scheme to implement a quantum phase
gate for trapped ions in thermal motion. In contrast to other
schemes \cite{cz,pcz,sm,sjj,gjm} only one standing wave laser
pulse is needed to implement a quantum controlled phase gate. Our
scheme does not require the rotating wave approximation. We also
show that the same scheme can be used to generate maximally
entangled states of $N$ trapped ions by using a
single standing wave laser pulse.\\
In the case of one trapped ion interacting with a standing wave
laser field, scheme have been proposed to generate and measure
vibrational state of a trapped ion\cite{a1}. Here we intend to
consider $N$ trapped ions with ground state $|g_i>$ and excited
state$|e_i>$ in a linear trap, which can interact with a
monochromatic standing wave of frequency $\omega_L$. We assume
that the ion-laser interaction strength characterized by Rabi
frequency $\Omega$ is the same for all ions and we also assume
that the position of the center of the all ions relative to a node
of the laser field is same, which is characterized by $\theta$.
The  total system is composed of the center of mass vibrational
freedom of the ion string and the internal electronic freedom of
the ions. The vibrational freedom is represented by a harmonic
oscillator with ladder operators $a^{\dagger}$ and $a$ and
frequency $\omega_a$ and the internal freedom is represented by
Pauli matrices $\sigma_{\alpha}^{(i)}$, $\alpha=x,y,z$. In a
frame, which rotates with the standing wave frequency $\omega_L$,
such a system can be described by the Hamiltonian
\begin{eqnarray}
H=\omega_a{a^{\dagger}a}+\sum_{i=1}^N\,\left[\frac{\Delta}{2}\sigma_{z}^{(i)}+\Omega\sin(kx+\theta)
\sigma_{x}^{(i)}\right]\,.\label{1}
\end{eqnarray}
Here $\sigma_{z}^{(i)}=|e_i><e_i|-|g_i><g_i|$ and
$\sigma_{x}^{(i)}=|e_i><g_i|+|g_i><e_i|$.  The parameter
$\Delta=\omega_0-\omega_L$ denotes the detuning between the
standing wave laser field and the atomic frequency $\omega_0$.
$kx=\eta(a+a^{\dagger})$ and $\eta$ is Lamb-Dicke parameter. In
the Lamb-Dicke limit, the Hamiltonian(\ref{1}) can be written as
\begin{eqnarray}
H=\omega_a{a^{\dagger}a}+\frac{\Delta}{2}{J_z}+\Omega[\eta(a+a^{\dagger})\cos\theta+\sin\theta]J_x\,.
\label{2}
\end{eqnarray}
Here we introduce collective spin operator
$J_{\alpha}=\sum_{i=1}^{N}\sigma_{\alpha}^{(i)}$, $\alpha=x,y,z$.
In the following, we consider a standing wave pulse, which is in
resonance with the carrier transition: $\Delta=0$. In this case,
the Hamiltonian (\ref{2}) simplifies to the form
\begin{eqnarray}
H=\omega_a{a^{\dagger}a}+\Omega[\eta(a+a^{\dagger})\cos\theta+\sin\theta]
J_x\,. \label{3}
\end{eqnarray}
The exact time evolution operator of this Hamiltonian (\ref{3}) is
\begin{eqnarray}
U(t)&=&\exp\left [iJ_x^2\left
(\frac{\eta^2\Omega^2\cos^2\theta{t}}{\omega_a}-\frac{\eta^2\Omega^2\cos^2\theta\sin\omega_a{t}}{\omega_a^2}\right
)\right ] \exp(-i\omega_a{a^{\dagger}a}t)\nonumber\\
&&\times\exp\left [-\frac{\eta\Omega\cos\theta}{\omega_a}J_x\left
(a^{\dagger}(e^{i\omega_a{t}}-1)-a(e^{-i\omega_a{t}}-1)\right
)\right ]\exp(-iJ_x\Omega\sin\theta{t})\,. \label{4}
\end{eqnarray}
 Note, that the rotating wave
approximation is not used. But our scheme uses a standing wave
field of a particular time length $\tau=2\pi/\omega_a$. The time
evolution operator reduces considerably
\begin{eqnarray}
U(\tau)=\exp\left
[\frac{2i\pi\eta^2\Omega^2\cos^2\theta}{\omega_a^2}J_x^2\right
]\exp\left [-\frac{2i\pi\Omega\sin\theta}{\omega_a}J_x\right ] \,.
\label{5}
\end{eqnarray}
It is easy to see, at the time $\tau$, two subsystem are
disentangled. The vibrational motion is returned to its original
state, be it the ground state or any vibrational excited state,
and we are left with an internal state evolution, which is
independent of the external vibrational state. In contrast to
other proposal, where the transformation (\ref{4}) was realized by
a sequence of laser pulses \cite{gjm,ms} our proposal
needs only a single standing wave pulse.\\
One possible application of the scheme is implementation of   a
quantum phase gate of two ions by using a single pulse laser. For
two trapped ions system, we choose the parameters $\eta$,
$\Omega$, $\theta$ and $\omega_a$ to satisfy
\begin{eqnarray}
\sin\theta=-\sqrt{\frac{\eta^2\pi}{\eta^2\pi+4}} \quad
\mbox{and}\quad
\frac{\omega_a}{\Omega}=\sqrt{\frac{64\eta^2}{\eta^2\pi^2+4\pi}}\,.
\label{6}
\end{eqnarray}
In this case, the time evolution operator becomes
\begin{eqnarray}
U(\tau)=\exp\left [\frac{i\pi}{8}(J_x^2+2J_x)\right ]\,.\label{7}
\end{eqnarray}
It is easy to check that this time evolution operator represents a
quantum phase gate
\begin{eqnarray}
|->_1|\pm>_2&\longrightarrow&|->_1|\pm>_2 \nonumber\\
|+>_1|->_2&\longrightarrow&|+>_1|->_2\nonumber\\
|+>_1|+>_2&\longrightarrow&-|+>_1|+>_2 \label{8}
\end{eqnarray}
in the basis states $|\pm>_i=\frac{1}{\sqrt{2}}(|e>_i\pm|g>_i)$.\\
In the following we will briefly demonstrate that it is possible
to prepare a multiparticle entanglement of $N$ trapped ions
\begin{eqnarray}
|\Psi>=\frac{1}{\sqrt{2}}[e^{i\varphi_g}|gg\cdots{g}>+e^{i\varphi_e}|ee\cdots{e}>]\,.
\label{9}
\end{eqnarray}
by using a single pulse laser, irrespective of N even or odd.
Quantum states of this kind were used to improve the frequency
standard. Several schemes \cite{bo,ms,sg} were proposed to
generate this kind of quantum states by a sequence of laser
pulses. Our proposal shows an improvement over these schemes,
since it requires only a single standing wave laser pulse. The
quantum state (\ref{9}) can be generated by this pulse, if the
system was initially prepared in the ground state $|ggg\cdots>$.
If $N$ is even, the parameter $\theta$ has to be set to zero. If
we choose this parameter to satisfy
$\frac{\Omega}{\omega_a}=\frac{1}{4\eta}$, the time evolution
operator will at the time $\tau=2\pi/\omega_a$ transform the
initial state with $U(\tau)=\exp\left [\frac{i\pi}{8}J_x^2\right
]$. In Ref.\cite{ms} such a kind of time evolution operator was
used to generate quantum states of the form (\ref{9}) with
$\varphi_g=-\frac{\pi}{4}$ and
$\varphi_e=\frac{\pi}{4}+\frac{N\pi}{2}$. In addition, it is also
pointed that for the time evolution operator (\ref{7}) can be used
to generate the maximally entangled state (\ref{9}) of odd $N$ trapped ions.\\
In summary, we proposed a novel scheme to implement a quantum
phase gate and entangle the electronic states of $N$ ions in a
very efficient way by a single standing wave laser pulse. This is
demonstrated in the case of the maximally entangled quantum state
(\ref{9}). The analysis of the system's dynamics, which we present
in this paper, doesn't use the rotation wave approximation. In
contrast to other schemes only the requirement of small Lamb-Dicke
parameters is needed. We calculate the exact time evolution and
set the time interval of the laser pulse exactly to the time of
one period of the collective ion vibration. This requirement helps
to fix the quantum number of the collective vibration. Since our
scheme is not dependent on the value of the vibration quanta, it
can be applied on ion systems in thermal motion. We pointed out
how this scheme makes use of the counter-rotating operator terms.
For the same purpose of implementing quantum phase gate and
generating GHZ state of N ions, other schemes
\cite{cz,pcz,sm,sjj,gjm} need three laser pulses or more.

\end{document}